\documentclass[conference]{IEEEtran}
\IEEEoverridecommandlockouts
\usepackage{cite}
\usepackage{amsmath,amssymb,amsfonts}
\usepackage[nolist]{acronym}
\usepackage{hyperref}
\usepackage{scalerel}
\usepackage{graphicx}
\usepackage{nicefrac}
\usepackage{xcolor}
\usepackage{circuitikz}

\usetikzlibrary{calc}
\usepackage{pgfplots}
\usepgfplotslibrary{groupplots}

\pgfplotsset{compat=1.15}

\makeatletter
\let\MYcaption\@makecaption
\makeatother

\usepackage[font=footnotesize]{subcaption}

\definecolor{espargosdarkgreen}{HTML}{227b3d}

\makeatletter
\let\@makecaption\MYcaption
\makeatother

\DeclareMathOperator*{\argmin}{arg\,min}

\usetikzlibrary{positioning}
\usetikzlibrary{svg.path}

\def\BibTeX{{\rm B\kern-.05em{\sc i\kern-.025em b}\kern-.08em
    T\kern-.1667em\lower.7ex\hbox{E}\kern-.125emX}}

\definecolor{orcidlogocol}{HTML}{A6CE39}
\tikzset{
  orcidlogo/.pic={
    \fill[orcidlogocol] svg{M256,128c0,70.7-57.3,128-128,128C57.3,256,0,198.7,0,128C0,57.3,57.3,0,128,0C198.7,0,256,57.3,256,128z};
    \fill[white] svg{M86.3,186.2H70.9V79.1h15.4v48.4V186.2z}
                 svg{M108.9,79.1h41.6c39.6,0,57,28.3,57,53.6c0,27.5-21.5,53.6-56.8,53.6h-41.8V79.1z M124.3,172.4h24.5c34.9,0,42.9-26.5,42.9-39.7c0-21.5-13.7-39.7-43.7-39.7h-23.7V172.4z}
                 svg{M88.7,56.8c0,5.5-4.5,10.1-10.1,10.1c-5.6,0-10.1-4.6-10.1-10.1c0-5.6,4.5-10.1,10.1-10.1C84.2,46.7,88.7,51.3,88.7,56.8z};
  }
}

\newcommand\orcidicon[1]{\href{https://orcid.org/#1}{\mbox{\scalerel*{
\begin{tikzpicture}[yscale=-1,transform shape]
\pic{orcidlogo};
\end{tikzpicture}
}{|}}}}

\begin{acronym}
 \acro{R2M}{raw 2\textsuperscript{nd} moment}
 \acro{CSI}{channel state information}
 \acro{UE}{user equipment}
 \acro{EM}{electromagnetic}
 \acro{UL}{uplink}
 \acro{BS}{base station}
 \acro{TDD}{time division duplex}
 \acro{FDD}{frequency division duplex}
 \acro{ECC}{error-correcting code}
 \acro{MLD}{maximum likelihood decoding}
 \acro{HDD}{hard decision decoding}
 \acro{IF}{intermediate frequency}
 \acro{RF}{radio frequency}
 \acro{SDD}{soft decision decoding}
 \acro{NND}{neural network decoding}
 \acro{CNN}{convolutional neural network}
 \acro{ML}{maximum likelihood}
 \acro{GPU}{graphical processing unit}
 \acro{BP}{belief propagation}
 \acro{LTE}{Long Term Evolution}
 \acro{BER}{bit error rate}
 \acro{SNR}{signal-to-noise-ratio}
 \acro{ReLU}{rectified linear unit}
 \acro{BPSK}{binary phase shift keying}
 \acro{QPSK}{quadrature phase shift keying}
 \acro{AWGN}{additive white Gaussian noise}
 \acro{MSE}{mean squared error}
 \acro{LLR}{log-likelihood ratio}
 \acro{MAP}{maximum a posteriori}
 \acro{NVE}{normalized validation error}
 \acro{BCE}{binary cross-entropy}
 \acro{CE}{cross-entropy}
 \acro{BLER}{block error rate}
 \acro{SQR}{signal-to-quantisation-noise-ratio}
 \acro{MIMO}{multiple-input multiple-output}
 \acro{mMIMO}{massive multiple-input multiple-output}
 \acro{OFDM}{orthogonal frequency division multiplex}
 \acro{RF}{radio frequency}
 \acro{LoS}{line of sight}
 \acro{NLoS}{non-line of sight}
 \acro{NMSE}{normalized mean squared error}
 \acro{CFO}{carrier frequency offset}
 \acro{SFO}{sampling frequency offset}
 \acro{IPS}{indoor positioning system}
 \acro{TRIPS}{time-reversal IPS}
 \acro{RSSI}{received signal strength indicator}
 \acro{MIMO}{multiple-input multiple-output}
 \acro{ENoB}{effective number of bits}
 \acro{AGC}{automatic gain control}
 \acro{ADC}{analog to digital converter}
 \acro{ADCs}{analog to digital converters}
 \acro{FB}{front bandpass}
 \acro{FPGA}{field programmable gate array}
 \acro{JSDM}{Joint Spatial Division and Multiplexing}
 \acro{NN}{neural network}
 \acro{IF}{intermediate frequency}
 \acro{LoS}{line-of-sight}
 \acro{NLoS}{non-line-of-sight}
 \acro{DSP}{digital signal processing}
 \acro{AFE}{analog front end}
 \acro{SQNR}{signal-to-quantisation-noise-ratio}
 \acro{SINR}{signal-to-interference-noise-ratio}
 \acro{ENoB}{effective number of bits}
 \acro{PCB}{printed circuit board}
 \acro{EVM}{error vector mangnitude}
 \acro{CDF}{cumulative distribution function}
 \acro{MRC}{maximum ratio combining}
 \acro{MRP}{maximum ratio precoding}
 \acro{MRT}{maximum ratio transmission}
 \acro{DeepL}{deep-learning}
 \acro{DL}{downlink}
 \acro{SISO}{single-input single-output}
 \acro{SGD}{stochastic gradient descent}
 \acro{CP}{cyclic prefix}
 \acro{MISO}{Multiple Input Single Output}
 \acro{LMMSE}{linear minimum mean square error}
 \acro{ZF}{zero forcing}
 \acro{USRP}{universal software radio peripheral}
 \acro{RNN}{recurrent neural network}
 \acro{GRU}{gated recurrent unit}
 \acro{LSTM}{long short-term memory}
 \acro{NTM}{neural turing machine}
 \acro{DNC}{differentiable neural computer}
 \acro{TCN}{temporal convolutional network}
 \acro{FCL}{fully connected layer}
 \acro{MANN}{memory augmented neural network}
 \acro{RNN}{recurrent neural network}
 \acro{DNN}{deep neural network}
 \acro{FIR}{finite impulse response}
 \acro{BPTT}{back-propagation through time}
 \acro{GAN}{generative adversarial network}
 \acro{ELU}{exponential linear unit}
 \acro{tanh}{hyperbolic tangent}
 \acro{BICM}{bit-interleaved coded modulation}
 \acro{OTA}{over-the-air}
 \acro{IM}{intensity modulation}
 \acro{DD}{direct detection}
 \acro{RL}{reinforcement learning}
 \acro{SDR}{software-defined radio}
 \acro{WGAN}{Wasserstein generative adversarial network}
 \acro{BMD}{bit-metric decoding}
 \acro{BMI}{bit-wise mutual information}
 \acro{LDPC}{low-density parity-check}
 \acro{IDD}{iterative demapping and decoding}
 \acro{JSD}{Jensen-Shannon divergence}
 \acro{MMSE}{minimum mean square error}
 \acro{FFT}{fast Fourier transform}
 \acro{DFT}{discrete Fourier transform}
 \acro{IFFT}{inverse fast Fourier transform}
 \acro{QAM}{quadrature amplitude modulation}
 \acro{EMD}{earth mover's distance}
 \acro{TDL}{tapped delay line}
 \acro{KL}{Kullback-Leibler}
 \acro{PRACH}{physical random access channel}
 \acro{URLLC}{ultra-reliable low-latency communication}
 \acro{ANOMA}{asynchronous non-orthogonal multiple access}
 \acro{FEC}{forward error correction}
 \acro{PAPR}{peak-to-average power ratio}
 \acro{APP}{a posteriori probability}
 \acro{COTS}{commercial off-the-shelf}
 \acro{PLL}{phase locked loop}
 \acro{STO}{sampling time offset}
 \acro{SFO}{sampling frequency offset}
 \acro{CFO}{carrier frequency offset}
 \acro{CPO}{carrier phase offset}
 \acro{CSI}{channel state information}
 \acro{GNSS}{global navigation satellite system}
 \acro{ELAA}{extremely large aperture array}
 \acro{UE}{user equipment}
 \acro{DICHASUS}{\underline{Di}stributed \underline{Cha}nnel \underline{S}ounder by \underline{U}niversity of \underline{S}tuttgart}
 \acro{JCaS}{Joint Communication and Sensing}
 \acro{CT}{continuity}
 \acro{TW}{trustworthiness}
 \acro{KS}{Kruskal's stress}
 \acro{PCA}{principal component analysis}
 \acro{AoA}{angle of arrival}
 \acro{ToA}{time of arrival}
 \acro{TDoA}{time difference of arrival}
 \acro{ToT}{time of transmission}
 \acro{RD}{Rajski's distance}
 \acro{ADP}{angle-delay profile}
 \acro{MPC}{multipath component}
 \acro{CIR}{channel impulse response}
 \acro{MAE}{mean absolute error}
 \acro{MDS}{multidimensional scaling}
 \acro{t-SNE}{t-distributed stochastic neighbor embedding}
 \acro{SM}{Sammon's mapping}
 \acro{CIRA}{channel impulse response amplitude}
 \acro{CS}{cosine similarity}
 \acro{FCF}{forward charting function}
 \acro{CMD}{correlation matrix distance}
 \acro{UWB}{ultra-wideband}
 \acro{MUSIC}{MUltiple SIgnal Classification}
 \acro{FBCM}{forward-backward correlation matrix}
 \acro{MDL}{minimum descriptive length}
 \acro{BFGS}{Broyden-Fletcher-Goldfarb-Shanno}
 \acro{UPA}{uniform planar array}
 \acro{RMS}{root mean square}
 \acro{DRMS}{distance root mean square}
 \acro{CEP}{circular error probable}
 \acro{R95}{95\textsuperscript{th} percentile radius}
 \acro{eCDF}{empirical cumulative distribution function}
 \acro{L-LTF}{legacy long training field}
 \acro{HT-LTF}{high throughput long training field}
 \acro{DOI}{Digital Object Identifier}
 \acro{PA}{Power Amplifier}
\end{acronym}

\definecolor{mittelblau}{RGB}{0, 126, 198}
\definecolor{violettblau}{cmyk}{0.9, 0.6, 0, 0}
\definecolor{rot}{RGB}{238, 28 35}
\definecolor{apfelgruen}{RGB}{140, 198, 62}
\definecolor{gelb}{RGB}{1, 221, 0}
\definecolor{orange}{RGB}{244, 111, 33}
\definecolor{pink}{RGB}{237, 0, 140}
\definecolor{lila}{RGB}{128, 10, 145}
\definecolor{hellgrau}{RGB}{224, 224, 224}
\definecolor{mittelgrau}{RGB}{128, 128, 128}
\definecolor{dunkelgrau}{RGB}{80,80,80}
\definecolor{anthrazit}{RGB}{19, 31, 31}

\makeatletter
\expandafter\def\csname pgf@anchor@rectangle@out A\endcsname{
	\pgf@process{\southwest}%
	\pgf@ya=0.25\pgf@y%
	\pgf@process{\northeast}%
	\pgf@y=0.75\pgf@y%
	\advance\pgf@y by\pgf@ya    }
\expandafter\def\csname pgf@anchor@rectangle@out B\endcsname{
	\pgf@process{\southwest}%
	\pgf@ya=0.75\pgf@y%
	\pgf@process{\northeast}%
	\pgf@y=0.25\pgf@y%
	\advance\pgf@y by\pgf@ya    } 
\expandafter\def\csname pgf@anchor@rectangle@in A\endcsname{
	\pgf@process{\southwest}%
	\pgf@ya=0.25\pgf@y%
	\pgf@process{\northeast}%
	\pgf@y=0.75\pgf@y%
	\advance\pgf@x by -15mm%
	\advance\pgf@y by\pgf@ya    }
\expandafter\def\csname pgf@anchor@rectangle@in B\endcsname{
	\pgf@process{\southwest}%
	\pgf@ya=0.75\pgf@y%
	\pgf@process{\northeast}%
	\pgf@y=0.25\pgf@y%
	\advance\pgf@x by -15mm%
	\advance\pgf@y by\pgf@ya    } 
\makeatother

\tikzset%
{
splitter/.style = {rectangle, draw, semithick, minimum height=5mm, minimum width=7.5mm,
							append after command={\pgfextra{\let\LN\tikzlastnode
	 \draw[thick] (\LN.west) -- ([xshift=-2mm] \LN.center)
							 -- ([xshift=-4mm] \LN.out A)
							 -- (\LN.out A)
  ([xshift=-2mm] \LN.center) -- ([xshift=-4mm] \LN.out B)
							 -- (\LN.out B);
	\node[left,font=\footnotesize] at (\LN.east) {#1};
												}}
						},
		 splitter/.default = {}
}

\tikzset%
{
combiner/.style = {rectangle, draw, semithick, minimum height=10mm, minimum width=15mm,
							append after command={\pgfextra{\let\LN\tikzlastnode
	 \draw[thick] (\LN.east) -- ([xshift=4mm] \LN.center)
							 -- ([xshift=8mm] \LN.in A)
							 -- (\LN.in A)
  ([xshift=4mm] \LN.center) -- ([xshift=8mm] \LN.in B)
							 -- (\LN.in B);
	\node[right,font=\footnotesize] at (\LN.west) {#1};
												}}
						},
	combiner/.default = {}
}

\newcommand{\ESPARGOS}[4] {
	\begin{scope}[shift = {#2}, scale = #3, rotate = #4, transform shape]
        \node (#1) at (0.375, 1.125) [minimum width = 1.5cm, minimum height = 3cm, fill = espargosdarkgreen, anchor = center] {};

        \node at (0, 0) [minimum width = 0.35cm, minimum height = 0.35cm, fill = white!95!black, rounded corners = 0.05cm] {};
        \node at (0.75, 0) [minimum width = 0.35cm, minimum height = 0.35cm, fill = white!95!black, rounded corners = 0.05cm] {};
        \node at (0, 0.75) [minimum width = 0.35cm, minimum height = 0.35cm, fill = white!95!black, rounded corners = 0.05cm] {};
        \node at (0.75, 0.75) [minimum width = 0.35cm, minimum height = 0.35cm, fill = white!95!black, rounded corners = 0.05cm] {};
        \node at (0, 1.5) [minimum width = 0.35cm, minimum height = 0.35cm, fill = white!95!black, rounded corners = 0.05cm] {};
        \node at (0.75, 1.5) [minimum width = 0.35cm, minimum height = 0.35cm, fill = white!95!black, rounded corners = 0.05cm] {};
        \node at (0, 2.25) [minimum width = 0.35cm, minimum height = 0.35cm, fill = white!95!black, rounded corners = 0.05cm] {};
        \node at (0.75, 2.25) [minimum width = 0.35cm, minimum height = 0.35cm, fill = white!95!black, rounded corners = 0.05cm] {};

        \node at (0, 0) [minimum width = 0.2cm, minimum height = 0.2cm, fill = white!70!black, inner sep = 0pt] {};
        \node at (0.75, 0) [minimum width = 0.2cm, minimum height = 0.2cm, fill = white!70!black, inner sep = 0pt] {};
        \node at (0, 0.75) [minimum width = 0.2cm, minimum height = 0.2cm, fill = white!70!black, inner sep = 0pt] {};
        \node at (0.75, 0.75) [minimum width = 0.2cm, minimum height = 0.2cm, fill = white!70!black, inner sep = 0pt] {};
        \node at (0, 1.5) [minimum width = 0.2cm, minimum height = 0.2cm, fill = white!70!black, inner sep = 0pt] {};
        \node at (0.75, 1.5) [minimum width = 0.2cm, minimum height = 0.2cm, fill = white!70!black, inner sep = 0pt] {};
        \node at (0, 2.25) [minimum width = 0.2cm, minimum height = 0.2cm, fill = white!70!black, inner sep = 0pt] {};
        \node at (0.75, 2.25) [minimum width = 0.2cm, minimum height = 0.2cm, fill = white!70!black, inner sep = 0pt] {};
        
        \node [rotate = 90, anchor = center] at (0.375, 1.125) { \includegraphics[width=1.5cm]{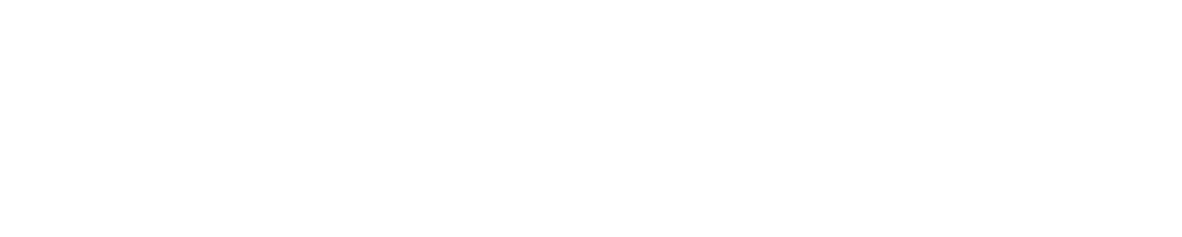} };
	\end{scope}
}

\begin{document}

\title{ESPARGOS: Phase-Coherent WiFi CSI Datasets for Wireless Sensing Research
\thanks{This work is supported by the German Federal Ministry of Education and Research (BMBF) within the project Open6GHub (grant no. 16KISK019).}}

\author{\IEEEauthorblockN{Florian Euchner\textsuperscript{\orcidicon{0000-0002-8090-1188}}, Stephan ten Brink\textsuperscript{\orcidicon{0000-0003-1502-2571}} \\}

\IEEEauthorblockA{
Institute of Telecommunications, Pfaffenwaldring 47, University of  Stuttgart, 70569 Stuttgart, Germany \\ \{euchner,tenbrink\}@inue.uni-stuttgart.de
}
}

\maketitle

\begin{abstract}
The use of WiFi signals to sense the physical environment is gaining popularity, with some common applications being motion detection and transmitter localization.
Standard-compliant WiFi provides a cost effective, easy and backward-compatible approach to Joint Communication and Sensing and enables a seamless transfer of results from experiments to practical applications.
However, most WiFi sensing research is conducted on channel state information (CSI) data from current-generation devices, which are usually not meant for sensing applications and thus lack sufficient spatial diversity or phase synchronization.
With ESPARGOS, we previously developed a phase-coherent, real-time capable many-antenna WiFi channel sounder specifically for wireless sensing.
We describe how we use ESPARGOS to capture large CSI datasets that we make publicly available.
The datasets are extensively documented and labeled, for example with information from reference positioning systems, enabling data-driven and machine learning-based research.
\end{abstract}


\section{Introduction}
Neural networks and other machine learning technologies are widely regarded as key enablers for next-generation wireless systems and have long been a focal point in telecommunications research.
In fields like computer vision and natural language processing, the availability of standardized training datasets has facilitated the development, performance evaluation, and comparison of machine learning algorithms.
However, in wireless research, particularly within the context of massive \ac{MIMO} systems, the availability of high-quality \ac{CSI} datasets remains limited \cite{bjornson2019massive}, forcing researchers to rely on channel models or simulated data.
To address this gap, we introduced DICHASUS \cite{dichasus2021}, a distributed massive MIMO channel sounder, which we regularly use to generate and publish highly calibrated \ac{CSI} data\footnote{DICHASUS datasets are available at \url{https://dichasus.inue.uni-stuttgart.de/}}.
These datasets have been applied across various fields, including differentiable ray tracing \cite{sionna-rt-calibration}, Channel Charting \cite{taner2024channel}, channel modeling, and other work related to \ac{JCaS} \cite{foliadis2024transfer}.
Channel sounders like DICHASUS are designed specifically to collect high-quality \ac{CSI}, and are typically neither real-time capable nor standards-compliant, often relying on \acp{SDR} at both ends of the link.

\begin{figure}
    \centering
    \includegraphics[width=0.55\columnwidth]{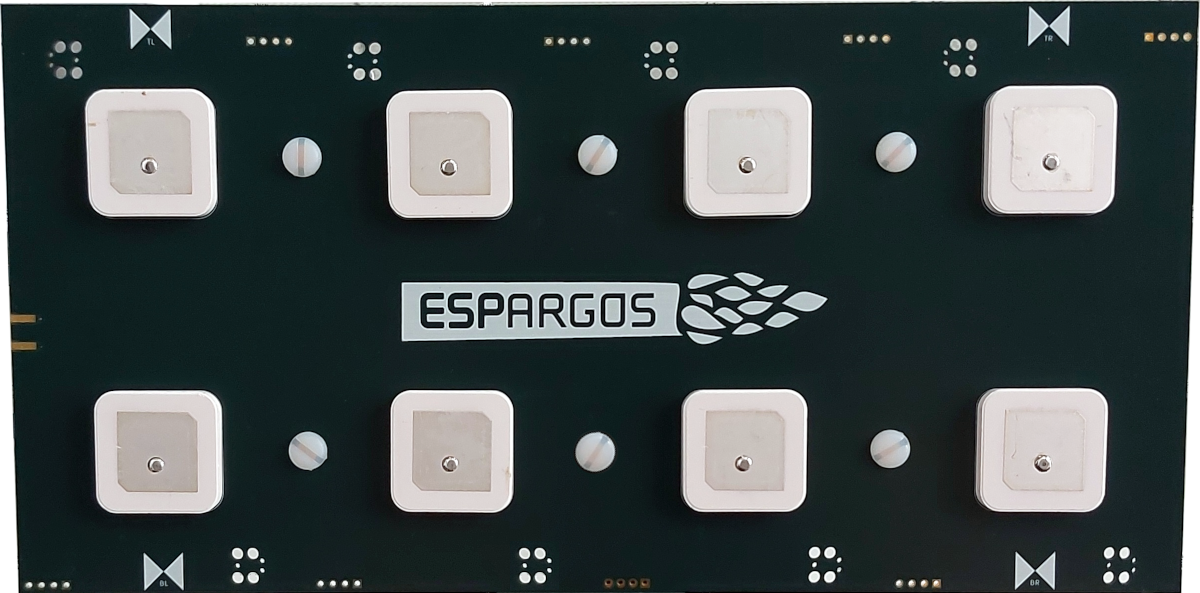}
    \caption{ESPARGOS antenna array with $2 \times 4$ patch antennas.}
    \vspace{-0.1cm}
    \label{fig:espargos-array-frontal}
    \vspace{-0.3cm}
\end{figure}

In a related line of research, it was observed that some commercially available WiFi chips offer \ac{CSI} dumping functionality, allowing \ac{CSI} to be extracted and processed \cite{Halperin_csitool} \cite{atheros_csitool}.
The use of WiFi has an obvious advantage: research outcomes can be easily transferred from theory to practical application.
However, most compatible devices offer limited spatial diversity (few antennas), may lack meaningful synchronization, and, more broadly, were not designed with sensing applications in mind.
To combine the spatial diversity of custom channel sounders with the flexibility and compatibility of commercial WiFi chip-based solutions, we developed ESPARGOS \cite{espargos}, a low-cost, WiFi-based channel sounder. Although ESPARGOS uses off-the-shelf WiFi chips, our circuit board design ensures phase synchronization across an arbitrary number of WiFi antennas.
Both hardware and software were designed with sensing applications in mind, facilitating the development of real-time applications without the need to tinker with WiFi drivers to obtain \ac{CSI}.
Even though ESPARGOS is real-time capable, the benefits of dataset-driven research still apply, most notably the comparability of results and the quick and easy access to datasets measured for environments and scenarios of interest.
For these reasons, we now also publish \ac{CSI} datasets measured with ESPARGOS\footnote{ESPARGOS datasets are available at \url{https://espargos.net/}}.
In the following, we summarize key features of the ESPARGOS system (Sec. \ref{sec:espargos}), document the collection and processing of WiFi \ac{CSI} datasets (Sec. \ref{sec:datasets}) and, using Channel Charting as an example, demonstrate how ESPARGOS datasets can accelerate the development of data-driven, real-time capable applications (Sec. \ref{sec:application}).
Finally, we conclude with an outlook that also draws a comparison to DICHASUS (Sec. \ref{sec:outlook}).

\section{Summary: ESPARGOS Hardware and Software}
\label{sec:espargos}

At its current stage of development, an ESPARGOS antenna array consists of eight antennas, each connected to an off-the-shelf WiFi-capable microcontroller (Espressif ESP32-S2), arranged into two rows of four antennas as shown in Fig. \ref{fig:espargos-array-frontal}.
Over a bus interface, measured CSI from all antennas is streamed to a central controller, which aggregates the measurements and forwards them over an Ethernet connection.
As thoroughly explained in \cite{espargos}, by clocking all microcontrollers from the same crystal oscillator and by distributing a phase reference signal to compensate for phase ambiguities introduced by the \acp{PLL}, the \ac{CSI} measurements are synchronized in phase.
ESPARGOS is usually operated as a passive sniffer, i.e., it listens for WiFi transmissions by other devices, but never transmits anything by itself.
This way, it can be easily integrated into existing WiFi deployments.
When collecting datasets for publication, however, we use a dedicated transmitter device.
This device was specially configured to continuously transmit very short WiFi packets, which ensures that up-to-date \ac{CSI} is available in regular intervals.

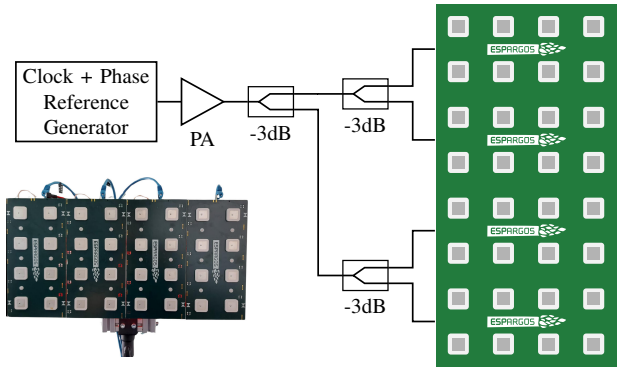
\begin{figure}
    \centering
    \scalebox{0.8}{
        \ctikzset{bipoles/thickness=1}
        \begin{circuitikz}[/tikz/circuitikz/bipoles/length=1cm, line width=0.8pt]
            \node (refgen) [draw, thick, anchor = east, align = center] at (0, -1.25) {Clock + Phase \\Reference\\Generator};

            \node (sref) [splitter, right = 1.5cm of refgen] {};
            \node (srefu) [splitter, right = 0.8cm of sref, yshift = 0.13cm] {};
            \node (srefl) [splitter, right = 0.8cm of sref, yshift = -2.89cm] {};
        
            \ESPARGOS{espargos0}{(5, 0)}{1}{-90};
            \ESPARGOS{espargos1}{(5, -1.51)}{1}{-90};
            \ESPARGOS{espargos2}{(5, -3.02)}{1}{-90};
            \ESPARGOS{espargos3}{(5, -4.53)}{1}{-90};
    
            \draw (refgen.east) to[amp, l_=PA] (sref.west);
            \draw [thick] (sref.out A) -- +(0.4, 0) |- (srefu.west);
            \draw [thick] (sref.out B) -- +(0.4, 0) |- (srefl.west);
            \draw [thick] (srefu.out A) -- +(0.4, 0) |- (espargos0.south);
            \draw [thick] (srefu.out B) -- +(0.4, 0) |- (espargos1.south);
            \draw [thick] (srefl.out A) -- +(0.4, 0) |- (espargos2.south);
            \draw [thick] (srefl.out B) -- +(0.4, 0) |- (espargos3.south);
            
            \node [below = 0.0cm of sref] {-3dB};
            \node [below = 0.0cm of srefu] {-3dB};
            \node [below = 0.0cm of srefl] {-3dB};
        
            \node at (-0.5, -4) {
                \includegraphics[width=4cm]{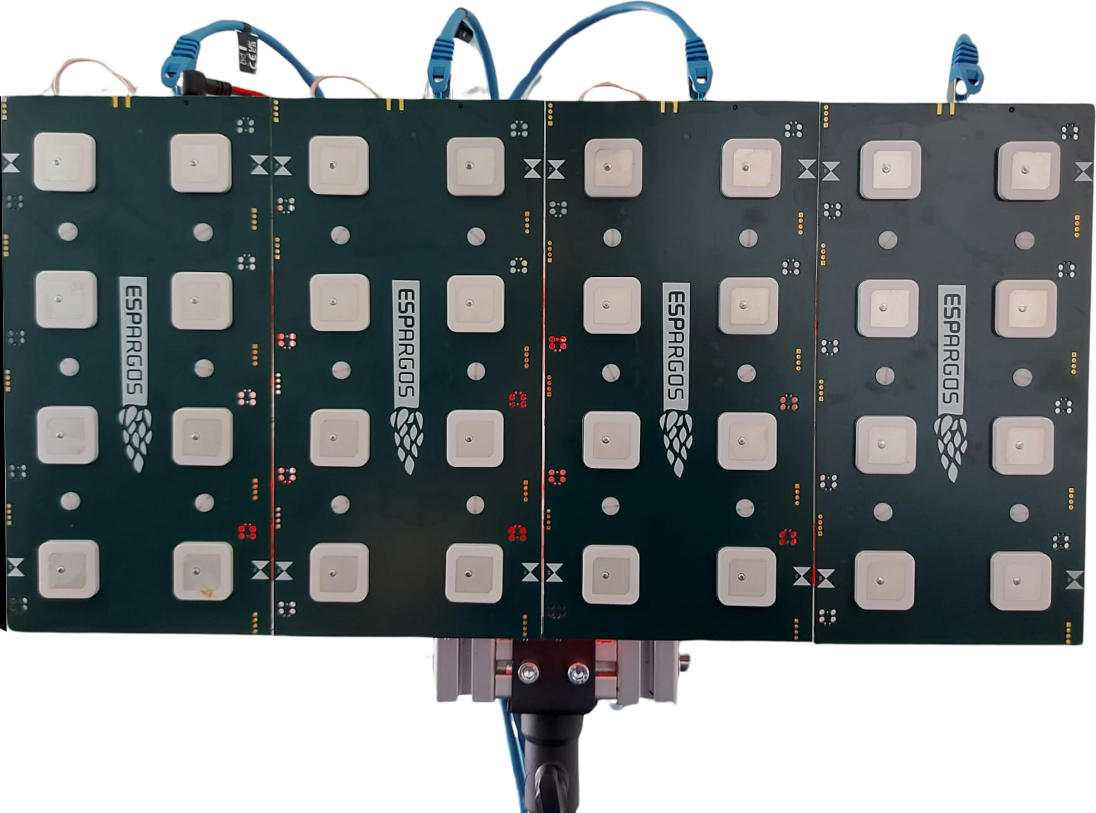}
            };
        \end{circuitikz}
    }
    \vspace{-0.1cm}
    \caption{With a common clock and phase reference signal, multiple ESPARGOS devices can be combined into a large phase-synchronous antenna array (schematic diagram and cropped photo of combined $4 \times 8$ array)}
    \label{fig:largearray}
    \vspace{-0.3cm}
\end{figure}

Multiple ESPARGOS boards can be combined into larger antenna arrays:
To this end, the clock signal (at $40\,\mathrm{MHz}$) and the phase reference signal (at around $2.4-2.5\,\mathrm{GHz}$), which are both generated by the central controller, are frequency-multiplexed so that they can be carried over a single coaxial cable.
The resulting combined reference signal is then amplified by a suitable \ac{PA} and distributed to all ESPARGOS boards over a cascade of power splitters, as shown in Fig. \ref{fig:largearray}.
If matching cable lengths are chosen and the power splitters offer sufficiently small phase unbalance, measured \ac{CSI} from all ESPARGOS boards is phase-coherent.
Otherwise, a constant phase offset may be introduced, which can, however, be accounted for in software.
To control multiple ESPARGOS devices from a central computer, a Python library called \emph{pyespargos}\footnote{Available at \url{https://github.com/ESPARGOS/pyespargos}} manages the board configuration and handles streaming \ac{CSI} in real-time.
A suite of demo applications shows how to use the library for applications like angle of arrival estimation.
To provide measurement datasets, a special application collects \ac{CSI} data provided by \emph{pyespargos} as well as metadata and writes everything to a file.

\section{Dataset Collection and Publication}
\label{sec:datasets}

\begin{figure}
    \centering
    \input{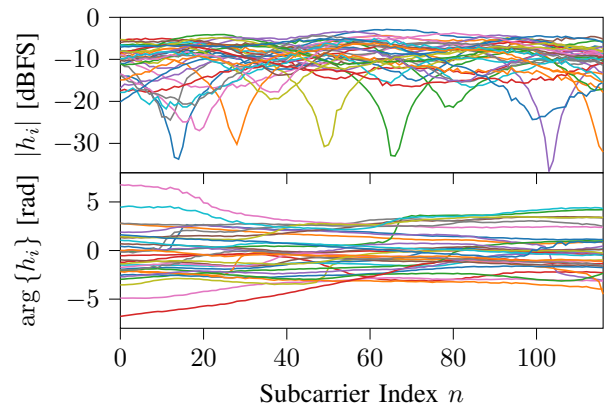}
    \vspace{-0.3cm}
    \caption{Visualization of $\mathbf { \bar H } \in \mathbb C^{B \times 2 \times 4 \times N_\mathrm{sub}}$ with $B = 4$ and $N_\mathrm{sub} = 117$: Channel coefficient amplitude and phase over subcarrier index $n = 0,\ldots,N_\mathrm{sub}-1$, each color representing one of the $B \times 2 \times 4$ antennas.}
    \label{fig:fdomain-csi}
    \vspace{-0.3cm}
\end{figure}

\begin{figure*}
    \centering
    \begin{subfigure}[b]{0.28\textwidth}
        \centering
        \includegraphics[height=3.6cm]{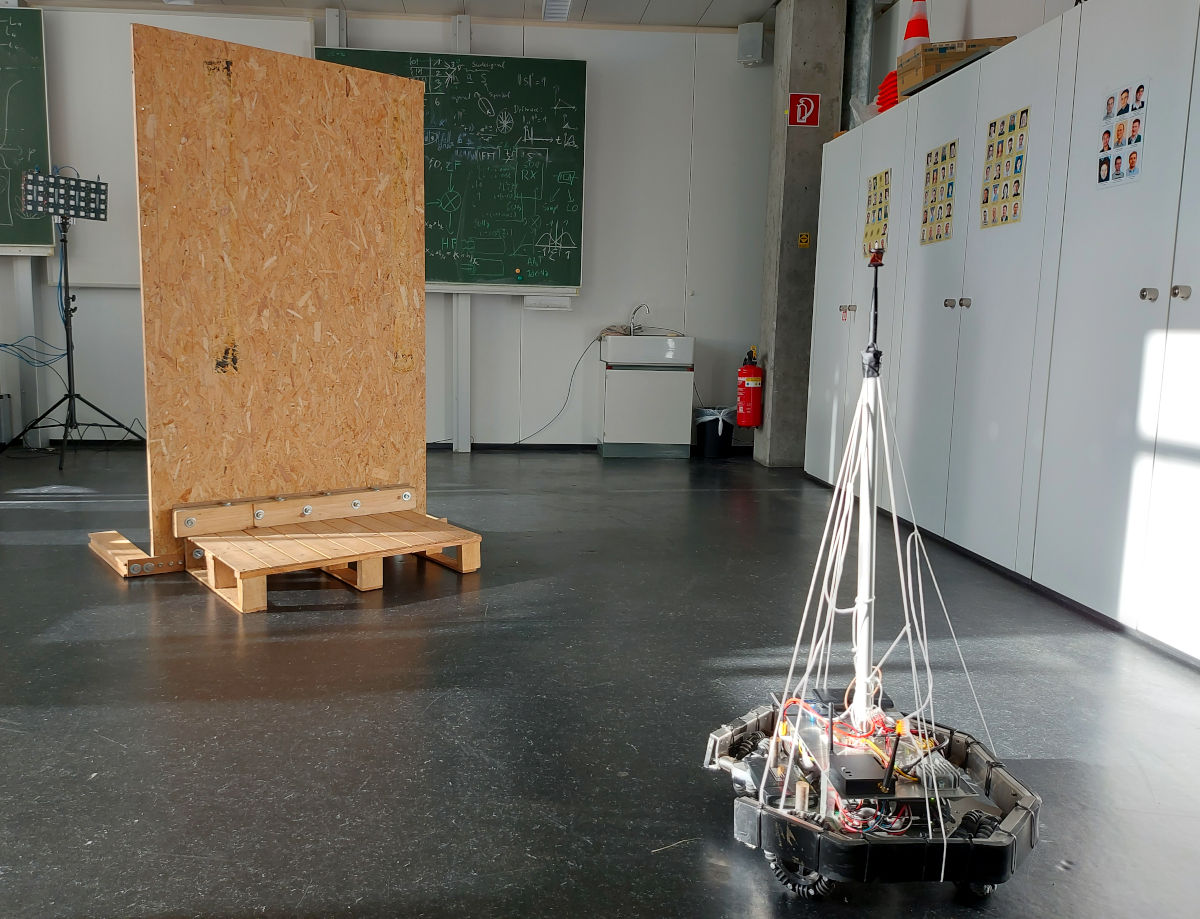}
        \vspace{0.7cm}
        \caption{}
        \label{fig:espargos-0002-photo}
    \end{subfigure}
     \begin{subfigure}[b]{0.35\textwidth}
        \centering
        \includegraphics[height=3.6cm]{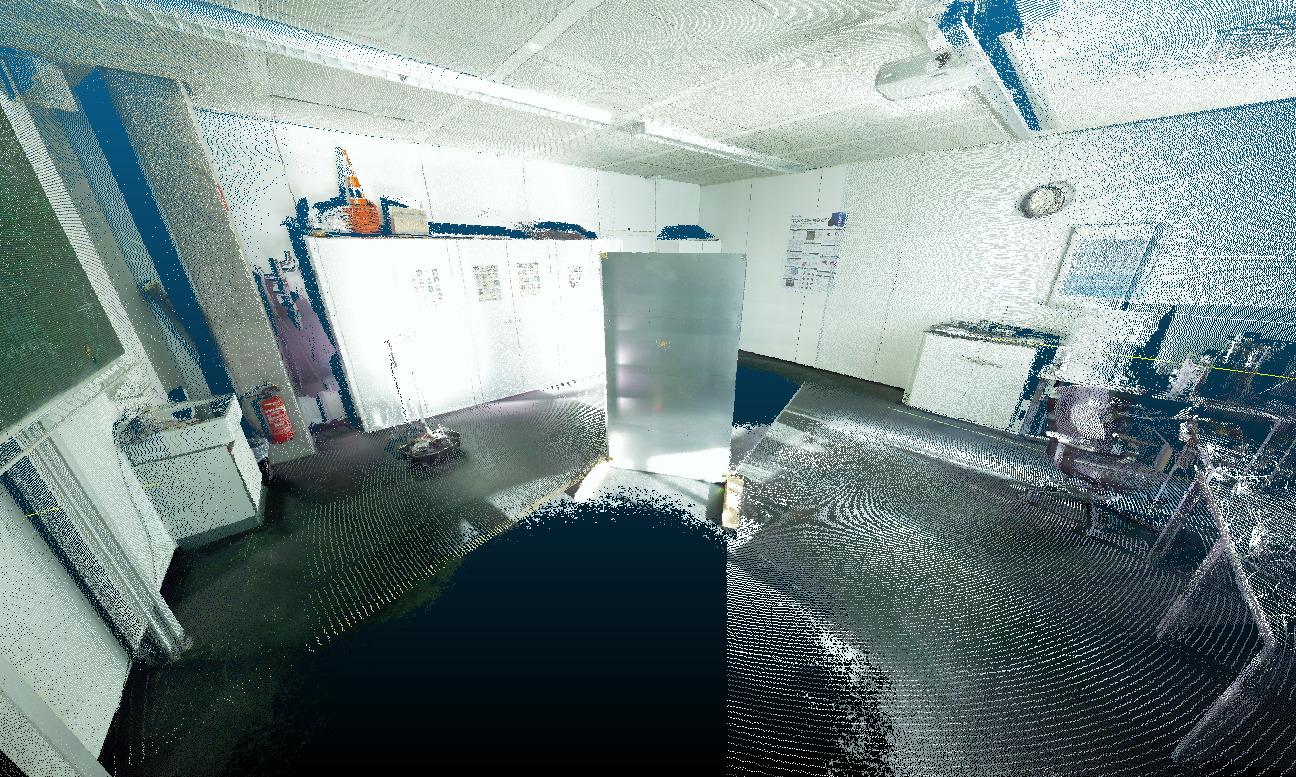}
        \vspace{0.7cm}
        \caption{}
        \label{fig:pointcloud}
    \end{subfigure}
   \begin{subfigure}[b]{0.30\textwidth}
        \centering
        \begin{tikzpicture}
            \begin{axis}[
                width=0.64\columnwidth,
                height=0.64\columnwidth,
                scale only axis,
                xmin=-6.5,
                xmax=-0.5,
                ymin=-1.5,
                ymax=5.5,
                xlabel = {Coordinate $x_1 ~ [\mathrm{m}]$},
                ylabel = {Coordinate $x_2 ~ [\mathrm{m}]$},
                ylabel shift = -4 pt,
                xlabel shift = -4 pt
            ]
                \addplot[thick,blue] graphics[xmin=-6,xmax=-0.5,ymin=-1.5,ymax=5] {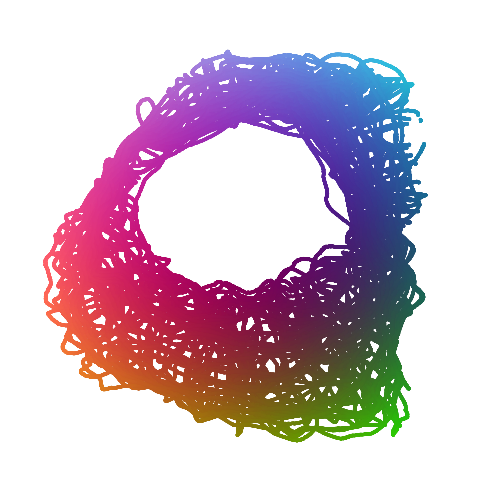};
                \draw [very thick] (axis cs:-2.855, 2.571) -- (axis cs:-3.712,1.974);
                \draw [green!50!black, very thick] (axis cs:-5.3399,4.5903) -- (axis cs:-5.0218,4.9378);
                
                \node (arrl) [fill = white, opacity = 0.5, text opacity = 1, align = center] at (axis cs: -2.5, 5) {\scriptsize ESPARGOS};
                \node (mwl) [fill = white, opacity = 0.5, text opacity = 1] at (axis cs: -3, 0) {\scriptsize Metal Wall};

                \draw [densely dotted] (mwl) -- (axis cs:-3.2835,2.2725);
                \draw [densely dotted] (arrl) -- (axis cs:-5.2223,4.7186);
            \end{axis}
        \end{tikzpicture}
        \vspace{-0.2cm}
        \caption{}
        \label{fig:topview}
    \end{subfigure}
    \vspace{-0.3cm}
    \caption{The exemplary dataset \textit{espargos-0002}: The figure shows (a) a photograph of the environment with the antenna array in the background, (b) a rendering of the 3D pointcloud and (c) a scatter plot (top view) of colorized ``ground truth'' positions of datapoints in $\mathcal S$, including antenna array and metal wall.}
    \label{fig:exemplary-dataset}
    \vspace{-0.4cm}
\end{figure*}
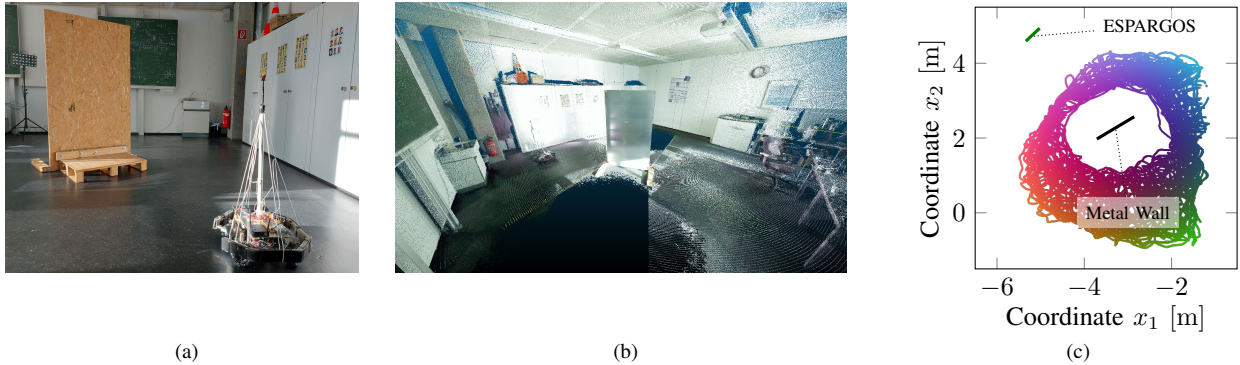

CSI datasets are collections of \ac{CSI} datapoints and associated metadata.
In the case of ESPARGOS, \ac{CSI} is estimated based on the \ac{HT-LTF} in each WiFi packet preamble.
A suitable WiFi packet that is captured by all ESPARGOS receivers in the system hence constitutes a \ac{CSI} datapoint, or, in other words, the channel coefficients contained within one \ac{CSI} datapoint correspond to the channel estimates produced by all receivers for the same WiFi packet.
The channel coefficients can be represented as a multidimensional array:
Since the receivers provide channel coefficients for all $N_\mathrm{sub}$ subcarriers, a system with $B$ ESPARGOS boards, each made up of $2$ rows of $4$ receivers, produces \ac{CSI} datapoints that contain complex-valued frequency-domain channel coefficients $\mathbf H \in \mathbb C^{B\times2\times4\times N_\mathrm{sub}}$.
Exemplary frequency-domain \ac{CSI} data $\mathbf { \bar H }$ is shown in Fig. \ref{fig:fdomain-csi}.
$\mathbf { \bar H }$ is the result of interpolating over 40 channel estimates $\mathbf H^{(l)}$ measured within an interval of $310\,\mathrm{ms}$.
The considered interval was arbitrarily selected from the \textit{espargos-0002} \cite{dataset-espargos-0002} \ac{CSI} dataset (see below).

In addition to channel coefficients, each receiver provides packet-specific \ac{RSSI} values, which are collected in $\mathbf P \in \mathbb R^{B\times2\times4}$.
To account for the effect of variable receiver gains, channel coefficients may be weighted by the corresponding \ac{RSSI}.
When collecting \ac{CSI} datasets with ESPARGOS, we make sure to also gather metadata such as the timestamp $t$ (in seconds) and the exact location of the transmitter $\mathbf x \in \mathbb R^3$ at the time of transmission.
To provide this data, which may be necessary ``ground truth'' for some machine learning applications, we use reference positioning systems such as a tachymeter total station (Leica MS60), which provides millimeter-level accuracy and high update rates.
We also make sure to account for any time offsets in the system, making sure to align transmission timestamps to the timestamps of the reference positioning system.
Combining \ac{CSI} and metadata, a dataset $\mathcal S$ containing a total of $L$ datapoints with indices $l$ can be written as

\[
    \mathcal S = \{ (\mathbf H^{(l)}, \mathbf P^{(l)}, \mathbf x^{(l)}, t^{(l)}) \}_{l = 1, \ldots, L}.
\]

In addition to datapoint-specific metadata, we also provide the exact location and orientation of all ESPARGOS arrays, photos of the environment and measurement setup and, for many datasets, even a 3-dimensional pointcloud scan of the environment, making sure that all relevant details of the measurement setup are documented.
All spatial information (antenna locations, pointcloud coordinates) is provided in the same right-handed cartesian coordinate system with meters as units, though the origin and orientation of the coordinate system are arbitrary.
We use a publication process that assigns a unique \ac{DOI} to each dataset, which, apart from the unique name, ensures unambiguous citability.

In the following, we consider a dataset called \textit{espargos-0002} \cite{dataset-espargos-0002}, which is introduced in Fig. \ref{fig:exemplary-dataset}:
As shown in Fig. \ref{fig:espargos-0002-photo}, the dataset was captured in an indoor lab room.
As previously described, $B = 4$ ESPARGOS antenna arrays are combined into a single phase-synchronous $4 \times 8$ array, and the WiFi transmitter is mounted to a robot, which is tracked by a total station.
By attaching the prism to the tip of the transmit antenna, which itself is on the tip of a pole, we ensure that the total station tracks the precise location of the antenna.
The robot is programmed to follow customized trajectories in the measurement area.
This way, as can be observed from Fig. \ref{fig:topview}, datapoints were captured in the measurement area with a high spatial density.
Due to the metal wall in the measurement area, which is drawn to scale in Fig. \ref{fig:topview} and also visible in Fig. \ref{fig:pointcloud}, many locations in the measurement area experience mostly non-line of sight propagation.

\section{Application Example: Channel Charting}
\label{sec:application}

\begin{figure}
    \centering
    \begin{subfigure}{0.49\columnwidth}
        \begin{tikzpicture}
            \begin{axis}[
                width=0.7\columnwidth,
                height=0.7\columnwidth,
                scale only axis,
                xmin=-9,
                xmax=9,
                ymin=-9,
                ymax=9,
                xlabel = {Coordinate $y_1$},
                ylabel = {Coordinate $y_2$},
                ylabel shift = -4 pt,
                xlabel shift = -4 pt,
                clip=true
            ]
                \addplot[thick,blue] graphics[xmin=-9,xmax=9,ymin=-9,ymax=9] {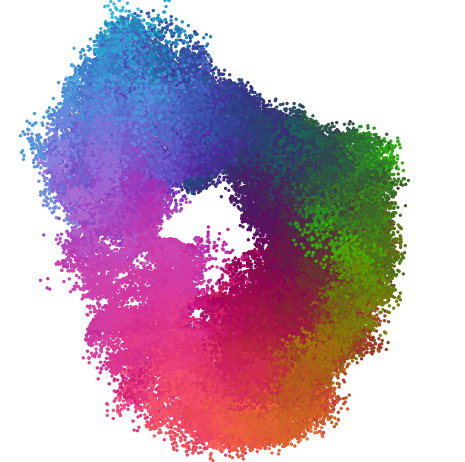};
                \draw [green!50!black, very thick] (axis cs:-5.3399,4.5903) -- (axis cs:-5.0218,4.9378);
            \end{axis}
        \end{tikzpicture}
        \vspace{-0.2cm}
        \caption{}
        \label{fig:channel-chart-before-transform}
    \end{subfigure}
    \begin{subfigure}{0.49\columnwidth}
        \begin{tikzpicture}
                \begin{axis}[
                width=0.7\columnwidth,
                height=0.7\columnwidth,
                    scale only axis,
                    xmin=-6.5,
                    xmax=-0.5,
                    ymin=-1.5,
                    ymax=5.5,
                    xlabel = {Coordinate $\hat x_1 ~ [\mathrm{m}]$},
                    ylabel = {Coordinate $\hat x_2 ~ [\mathrm{m}]$},
                    ylabel shift = -4 pt,
                    xlabel shift = -4 pt,
                    clip=true
                ]
                    \addplot[thick,blue] graphics[xmin=-6,xmax=-0.5,ymin=-1.5,ymax=5] {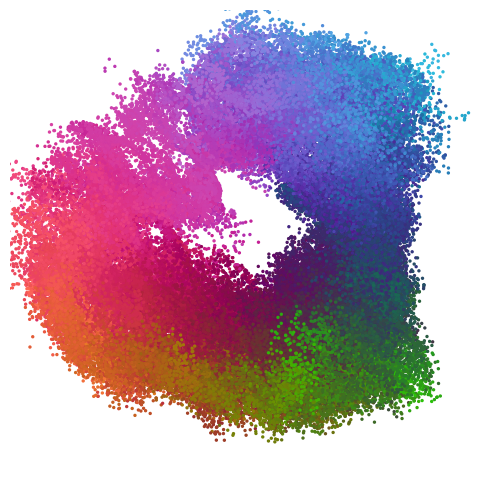};
                    \draw [green!50!black, very thick] (axis cs:-5.3399,4.5903) -- (axis cs:-5.0218,4.9378);
                \end{axis}
        \end{tikzpicture}
        \vspace{-0.2cm}
        \caption{}
        \label{fig:channel-chart-after-transform}
    \end{subfigure}
    \vspace{-0.3cm}
    \caption{Learned channel chart (a) before and (b) after optimal affine transformation, datapoint colorization is preserved from Fig. \ref{fig:topview}}
    \label{fig:channel-chart}
    \vspace{-0.4cm}
\end{figure}

As one use case of ESPARGOS \ac{CSI} datasets for research purposes, we consider the task of \ac{CSI}-based indoor localization.
From the earlier description of the \textit{espargos-0002} dataset, it is clear that there are no simple model-based approaches for indoor localization for the given scenario:
While it is possible to determine an angle of arrival for datapoints with a \ac{LoS} path between transmitter and receiver array, triangulation is not applicable with only one antenna array.
Time of arrival-based multilateriation is impossible because there is no time synchronization between transmitter and receiver.
Due to the dominance of multipath propagation in the indoor environment, it is also not reasonable to use \ac{RSSI} information to estimate the distance between transmitter and receiver based on path loss.
Despite these difficult circumstances, a data-driven, self-supervised approach for indoor localization like Channel Charting \cite{studer_cc} can at least provide relative position estimates without requiring reference positions.
In contrast to model-based approaches like triangulation or multilateration, which assume a \ac{LoS} channel, Channel Charting makes no such assumption about the propagation environment, but instead relies on similarity relationships between measured \ac{CSI} samples.
Since \ac{CSI} is so high-dimensional, it is commonly assumed that every possible location of the transmitter in physical space manifests itself in a unique representation in \ac{CSI} space, and that datapoints with transmitter locations which are close in physical space are also close in \ac{CSI} space.
Since the set of all possible locations of the transmitter is low-dimensional (in this case, two-dimensional) and the space of all possible channel realizations is high-dimensional, the problem can be understood as identifying and ``un-folding'' a low-dimensional manifold in high-dimensional \ac{CSI} space.
In that sense, Channel Charting is a dimensionality reduction techniques that finds a mapping $\mathcal C_\theta: \mathbb C^{B \times 2 \times 4 \times N_\mathrm{sub}} \to \mathbb R^2$, called \ac{FCF}, that maps high-dimensional CSI $\mathbf H \in \mathbb C^{B \times 2 \times 4 \times N_\mathrm{sub}}$ to coordinates $\mathbf y \in \mathbb R^2$ in a low-dimensional space called the channel chart.
This channel chart can be interpreted as a \emph{map} of the environment and while it may be distorted compared to the physical reality, it should at least provide meaningful information about relative positions.
Often, channel chart coordinates can be mapped to coordinates in physical space in a meaningful way by finding an affine transformation between these coordinate frames.

A detailed description of the Channel Charting algorithm is outside the scope of this work, but the Triplet Neural Network-based approach \cite{ferrand2020triplet} taken here is similar to what we describe in one of our previous publications \cite{euchner2022improving}, except for the different underlying dataset (now ESPARGOS and no longer DICHASUS) and tweaked hyperparameters
\footnote{Source code for Channel Charting with ESPARGOS datasets is available at \url{https://github.com/Jeija/ESPARGOS-WiFi-ChannelCharting}}.
For the following results, we use a subset of \emph{espargos-0002}, containing $L = 569190$ datapoints, though good results are also achievable with significantly fewer datapoints.
The resulting channel chart, shown in Fig. \ref{fig:channel-chart-before-transform}, is obtained by applying the learned \ac{FCF} to all datapoints in the dataset.
To facilitate a quick visual inspection, the colors assigned to the datapoints based on their reference positions (see Fig. \ref{fig:topview}) are preserved in the channel chart.
If the color gradient is reproduced in the channel chart, as is the case here, this indicates that the \ac{FCF} has managed to recover the global geometry of the environment.
While some areas of the channel chart are somewhat distorted, the overall ring shape remains clearly visible.
Obviously, the channel chart is rotated, scaled and also translated relative to the physical coordinate frame.
We find the optimal affine coordinate transform $T_\mathrm{c}(\mathbf y) = \hat {\mathbf A} \mathbf y + \hat {\mathbf b}$ from the channel chart's coordinate frame to physical coordinates by solving the least squares problem
\[
    (\mathbf{\hat A}, \mathbf{\hat b}) = \argmin\limits_{(\mathbf{A}, \mathbf{b})} \sum_{l = 1}^L \lVert\mathbf{A} {\mathbf y }^{(l)} + \mathbf b - \mathbf x^{(l)} \rVert_2^2,
\]
where $\mathbf y^{(l)} = \mathcal C_\theta(\mathbf H^{(l)})$ denotes the channel chart coordinates produced by the \ac{FCF} for the datapoint with index $l$, and $\mathbf x^{(l)}$ is the corresponding two-dimensional ``ground truth'' position from the dataset.
After $T_\mathrm{c}$ is applied to the channel chart, the resulting points ${ \mathbf {\hat x }^{(l)} } = T_\mathrm{c} \left(\mathcal C_\Theta(\mathbf H^{(l)})\right)$ are now in the correct physical coordinate frame as shown in Fig. \ref{fig:channel-chart-after-transform}.

Subjectively, the ``ground truth'' coordinates in Fig. \ref{fig:topview} and the transformed channel chart in Fig. \ref{fig:channel-chart-after-transform} look similar.
A more objective performance assessment is conducted in Tab. \ref{tab:cc-performance-metrics}, which evaluates the channel chart based on performance metrics which are commonly used in Channel Charting literature.
Compared to previous work on Channel Charting, the observed performance is acceptable, but leaves room for improvement.

\begin{table}
    \centering
    \caption{Performance Metrics for Channel Charting. Metrics marked with (*) were evaluated after the optimal affine transform $T_\mathrm{c}$.}
    \vspace{-0.2cm}
    \begin{tabular}{c | c | c | c | c}
        \textbf{CT} & \textbf{TW} & \textbf{KS} & \textbf{MAE}* & \textbf{CEP}* \\[0.01cm] \hline
        0.96 & 0.96 & 0.20 & $0.44\,\mathrm m$ & $0.42\,\mathrm m$
    \end{tabular}
    \vspace{0.1cm}

    \textbf{CT} = Continuity, \textbf{TW} = Trustworthiness, \textbf{KS} = Kruskal's Stress, \newline
    \textbf{MAE} = Mean Abs. Err., \textbf{CEP} = Circular Error Probable, as defined in \cite{asilomar2023}
    \label{tab:cc-performance-metrics}
    \vspace{-0.3cm}
\end{table}

\section{Conclusion and Outlook}
\label{sec:outlook}

With ESPARGOS, we developed a real-time capable phase-coherent WiFi channel sounder.
Owing to the benefits of dataset-driven research, we now also publish datasets containing \ac{CSI} measured by ESPARGOS, alongside relevant metadata.
Compared to data generated by our custom channel sounder DICHASUS (and most other \ac{SDR}-based channel sounders), \ac{CSI} measured by ESPARGOS is considerably more noisy and may exhibit additional impairments, as we lack control over some aspects of the WiFi chip.
The advantages of the WiFi-based approach are the much lower cost, backwards-compatibility and real-time capability, which partially makes up for the lower data quality.
The successful training of a \ac{FCF} for Channel Charting proves that the \ac{CSI} quality is at least sufficient for this type of application.

Since ESPARGOS is designed to be easy-to-use, the possibility of third-party dataset contributions is opened up.
This is in contrast to bespoke channel sounders, which usually require experienced operators.
While datasets and some application examples are publicly accessible already, we plan to continue to publish additional software components over time.
Applications developed with the help of measured datasets can be practically tested and demonstrated thanks to the real-time capability of ESPARGOS.
For example, we developed a real-time demonstrator for Channel Charting internally, and we plan to publish more details on this demonstrator soon.
Most importantly, we invite researchers to work with the publicly available ESPARGOS datasets.

\bibliographystyle{IEEEtran}
\bibliography{IEEEabrv,references}

\end{document}